\documentclass[prb,10pt,showpacs,superscriptaddress]{revtex4}

\usepackage[dvips]{epsfig}
\usepackage{subfigure}

\usepackage{amsfonts}
\usepackage{amsmath}
\usepackage{amssymb}
\usepackage{pslatex}
\usepackage{color}
\usepackage{hhline}
\usepackage{threeparttable}
\usepackage{supertabular}
\usepackage{multirow}
\usepackage{array}
\usepackage{tabularx}
\usepackage{rotating}
\usepackage{color}

\newcommand{\vecb}[1]{\mathbf{#1}}

\newcolumntype{Y}{>{\centering\arraybackslash}X}

\begin{document}

\bibliographystyle{apsrev}

\title{Design of mid-IR and THz quantum cascade laser cavities with complete TM photonic bandgap}

\author{Michael Bahriz}
\affiliation{Institut d'Electronique Fondamentale, Bat. 220, Universite Paris-Sud, 91405 Orsay, France}
\author{Orion Crisafulli}
\affiliation{Thomas J. Watson, Sr., Laboratory of Applied Physics, California Institute of Technology, Pasadena, California 91125}
\author{Virginie Moreau}
\author{Raffaele Colombelli}
\email{colombel@ief.u-psud.fr}
\affiliation{Institut d'Electronique Fondamentale, Bat. 220, Universite Paris-Sud, 91405 Orsay, France}

\author{Oskar Painter}
\email{opainter@caltech.edu}
\homepage{http://copilot.caltech.edu}
\affiliation{Thomas J. Watson, Sr., Laboratory of Applied Physics, California Institute of Technology, Pasadena, California 91125}
\date{\today}
 
\begin{abstract}
We present the design of mid-infrared and THz quantum cascade laser cavities formed from planar photonic crystals with a complete in-plane photonic bandgap. The design is based on a honeycomb lattice, and achieves a full in-plane photonic gap for transverse-magnetic polarized light while preserving a connected pattern for efficient electrical injection. Candidate defects modes for lasing are identified. This lattice is then used as a model system to demonstrate a novel effect: under certain conditions - that are typically satisfied in the THz range - a complete photonic gap can be obtained by the sole patterning of the top metal contact. This possibility greatly reduces the required fabrication complexity and avoids potential damage of the semiconductor active region.
\end{abstract} 

\maketitle

\setcounter{page}{1}

\section{Introduction}
Quantum cascade (QC) lasers are semiconductor laser sources based on intersubband (ISB) transitions in multiple quantum
well systems \cite{gmachl-1533}. Their emission wavelength can be tuned across the mid-infrared
\cite{qc-mid-ir1,qc-mid-ir2} (mid-IR, 5 $\mu$m $<$ $\lambda$ $<$ 24 $\mu$m) and THz \cite{QC-thz} (65 $\mu$m $<$
$\lambda$ $<$ 200 $\mu$m) ranges of the electromagnetic spectrum. In the mid-IR, where they are becoming the
semiconductor source of choice thanks to their good performance, the potential applications are chemical sensing,
spectroscopy and free-space optical communications. In the THz range the most promising application is the imaging of
concealed objects. Biological materials, semiconductor chip packaging and clothing are all THz-transparent (while they
are opaque at shorter wavelengths), making THz imaging useful for security and medical applications.

Most of the activity in the QC laser field has concentrated on edge-emitting lasers due to the intrinsic
transverse-magnetic (TM) polarization of ISB transitions, and corresponding difficulty in implementing vertical-cavity
surface emitting lasers (VCSELs).  Surface emission has been obtained, rather, by integrating second-order gratings on
edge emitting devices \cite{hofstetter-3769,schrenk-2086} or by replacing the standard Fabry-Perot cavity with a
photonic crystal (PC) resonator \cite{Colombelli03,kartik_apl04}.  The latter solution, the application of photonic
crystal technology to QC lasers \cite{joann-book,vurgaftman03}, is particularly appealing because of the flexibility
that it allows the designer.  Two-dimensional (2D) photonic crystals can be used to create localized microcavity laser
sources that can be built as two-dimensional arrays on a single chip\cite{PainterPRBSym03,PainterJOSAB99} or for
large-area, high-power, single-mode surface emitting laser sources\cite{imada02}.

Of particular importance in determining the properties of any photonic crystal structure is the effective refractive
index contrast attainable.  The higher the index contrast, the stronger the optical dispersion of the photonic bands and
the greater the ability to localize, diffract, and reflect light within the photonic lattice.  A high index contrast is
therefore crucial for device miniaturization.  In planar waveguide devices, the presence of a full optical band gap in
two-dimensions is beneficial (although not completely necessary) in forming ultra-low-volume laser cavities with minimal
optical loss.  As is discussed in Ref.\cite{meade1993}, for 2D photonic crystals, TM optical band gaps are favored in a
lattice of isolated high-$\epsilon$ regions. Unfortunately, this configuration is incompatible with an electrical
injection device due to its non-connected nature and an alternate approach is required.

In this paper we study the use of a connected honeycomb lattice for creating 2D photonic crystal QC laser structures.
We begin in Section \ref{sec:2D} with a review of the properties of this 2D lattice \emph{via} a planewave expansion
(PWE) analysis. We show that a full 2D optical bandgap for TM polarization can be obtained in this connected lattice,
and study the localized resonant modes that form around a simple point defect in the lattice.  In Sections
\ref{sec:3D-mir} and \ref{sec:3D-thz} full three-dimensional (3D) finite-difference time-domain (FDTD) simulations are
used to analyze the properties of the honeycomb lattice in two different vertical waveguide structures.  Section
\ref{sec:3D-mir} deals with mid-IR PC QC lasers. The real active-region and waveguide structure of a mid-IR
surface-plasmon (SP) QC laser is considered.  The high-index contrast in this waveguide structure is obtained by air
holes that penetrate deeply into the semiconductor layers.  Section \ref{sec:3D-thz} focuses on THz PC QC laser
waveguide structures. In the metal-metal waveguides of THz lasers we find that the effective index contrast, and
therefore the photonic bandstructure, is strongly dependent on the waveguide thickness\footnote{A similar phenomenon
  occurs in guided membrane PC structures, where it is known that the extent of the photonic gap depends on the membrane
  thickness \cite{johnson99}.  However, in the dielectric membrane structures, beyond a critical membrane thickness
  further reduction in thickness does not increase the bandgap due to a loss of mode localization in the dielectric
  membrane.  The double-metal waveguide structure does not suffer from such a loss of confinement.}. In particular,
below a certain critical waveguide thickness, a full photonic gap can be induced by the sole patterning of the metal
layers.  This novel effect could be useful for the development of a variety of THz lasers, including PC surface-emitting
lasers\cite{Schubert06}, due to the simple fabrication requirements.

\section{Two-dimensional analysis: planewave expansion}
\label{sec:2D}

\subsection{The planewave expansion method}
The calculations of this section are based on a planewave expansion (PWE) method. It is a frequency-domain method that allows extraction of the Bloch fields and frequencies by direct diagonalization. The calculations were performed using the MIT Photonic Bands (MPB) simulation tool\cite{Johnson2001}.  With the PWE method it is possible to compute the bandstructures and electromagnetic modes of perfectly periodic dielectric structures. Calculation of defect modes will be tackled in section \ref{sec:supercell} using a super-cell approach.

\subsection{Photonic crystal structure and TM gap}

\begin{figure}[htb]
\centering
\includegraphics[width=0.6\linewidth]{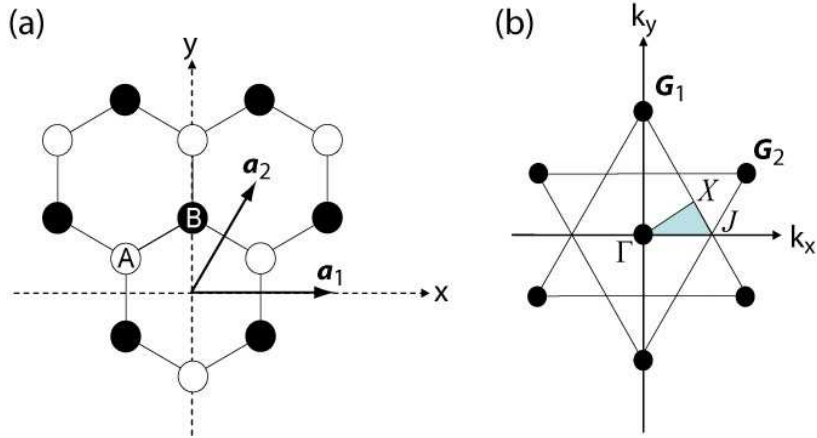}
\caption{Structure of the honeycomb lattice . (a) Two-dimensional honeycomb lattice with two "dielectric atoms'' ($A$ and $B$) per unit cell. $\vecb{a}_1=(a,0)$ and $\vecb{a}_2=(a/2,\sqrt{3}a/2)$ are the principal lattice vectors ($a=|\vecb{a}_1|=|\vecb{a}_2|$), $r$ is the hole radius. (b) Two-dimensional reciprocal space for the honeycomb lattice.  The reduced Brillouin zone is shaded in blue.  Reciprocal lattice vectors are superpositions of $\vecb{G}_{1}=(0,4\pi/\sqrt{3}a)$ and $\vecb{G}_{2}=(2\pi/a,2\pi/\sqrt{3}a)$.} 
\label{fig:PWEstruct}
\end{figure}

The 2D photonic crystal considered in this work is that of a honeycomb lattice, consisting of a hexagonal Bravais lattice with a basis of two air holes per unit cell (see Fig. \ref{fig:PWEstruct}(a)). The hole radius is $r$ and the lattice constant is $a$. The reciprocal lattice is also hexagonal, with $\Gamma$, $X$ and $J$ the high symmetry points of the lattice (see Fig. \ref{fig:PWEstruct}(b)).  In the 2D simulations of this section the relative dielectric constant of the dielectric background is taken as $\epsilon = 11.22$ ($n=3.35$), corresponding to the index of refraction of most III-V semiconductor materials at frequencies below the energy gap.

\begin{figure}[htb]
\centering
\includegraphics[width=0.75\linewidth]{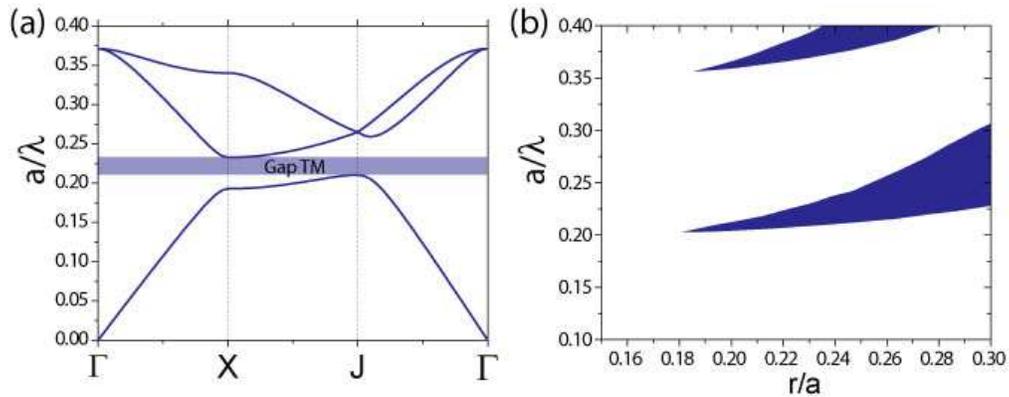}
\caption{Photonic band calculations for a 2D honeycomb lattice of air holes in a dielectric lattice ($\epsilon = 11.22$).  (a) Photonic bandstructure for TM polarization with $r/a$ = 0.234.  (b) Complete band gap (shaded blue region) as a function of $r/a$. A full TM gap exists only for $r/a > 0.18$} 
\label{fig:PWEgapmap}
\end{figure}

Fig. \ref{fig:PWEgapmap}(a) shows the photonic bandstructure for a lattice with $r/a=0.234$. For this $r/a$ ratio a full TM gap is present, centered around $a/\lambda \approx 0.225$, with a width approximately $12\%$ of the central gap frequency. The existence of a gap, however, depends strongly on $r/a$.  In Fig.\ref{fig:PWEgapmap}(b) we plot the gap-map for this lattice, which clearly shows that only above a critical value of circular hole size ($r/a \geq 0.18$) can a TM photonic gap be obtained. This is a common behavior of lattices of air holes; the larger the holes, the closer the lattice is to a system of isolated high-$\epsilon$ regions, and the larger the TM photonic bandgap\cite{joann-book}.  The advantage of the honeycomb lattice is that the TM gap opens when the hole size and porosity of the lattice are still reasonably small, allowing for more efficient electrical injection.    

\subsection{2D defect design: supercell method}
\label{sec:supercell}

\begin{figure}[htb]
\centering
\includegraphics[width=0.8\linewidth]{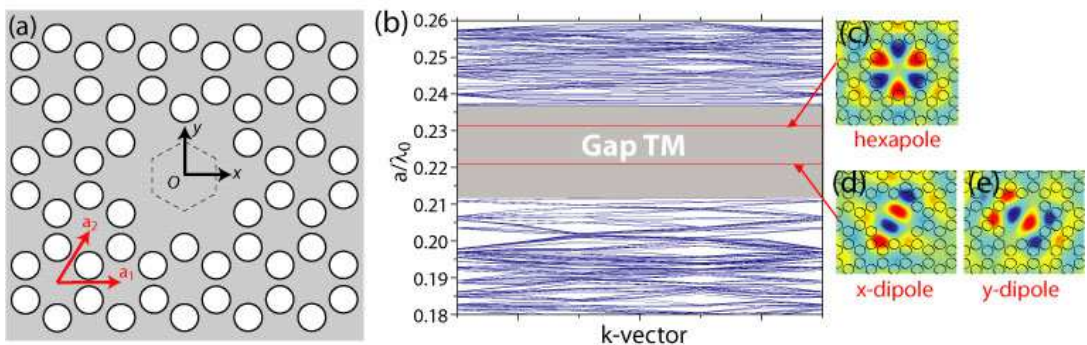}
\caption{(a) Illustration of the 2D honeycomb lattice defect cavity supercell (background dielectric material shown as grey, air holes shown as white).  The principle lattice vectors are $\vecb{a}_{1}$, $\vecb{a}_{2}$, where $|\vecb{a}_{1}| = |\vecb{a}_{2}| = a$.  The defect region consists of removal of the central hexagon of air holes (central hexagon shown as a dashed line).  (b) Folded TM photonic bandstructure for a 20-period supercell of the defect cavity shown in (a) with $r/a=0.24$.  The blue-lines correspond to the folded bandstructure of the honeycomb lattice.  The red lines are the defect frequency levels.  Localized defect modes: (c) hexapole and (d-e) dipole-like modes.
}
\label{fig:PWEbands}
\end{figure}

In this section we study, in 2D, the localized resonances that form around a point-like defect of the honeycomb lattice.  The defect that we consider is obtained by removing a hexagon of six holes, as shown in Fig. \ref{fig:PWEbands}(a).  The defect modes are calculated with a supercell method using the same PWE solver. The presence of the defect cavity breaks the periodicity of the lattice, requiring the creation of a supercell over which the structure is assumed periodic.  This supercell contains the defect cavity and is tiled in space. The Wigner-Seitz cell of the reciprocal lattice of the supercell will be correspondingly smaller than that of the underlying photonic lattice, resulting in a folding of the honeycomb lattice photonic bands (see Fig. \ref{fig:PWEbands}(b)).  The localized defect states of the supercell appear as bands with almost no dispersion, lying inside the photonic bandgap of the honeycomb lattice.   

A PWE calculated bandstructure of the supercell defect structure with $r/a = 0.24$ is shown in Fig. \ref{fig:PWEbands}(b).  In these calculations, a supercell consisting of 20 periods of the honeycomb lattice was used in order to obtain well localized resonant modes.  Three defect modes lie within the complete band gap: two degenerate dipole-like modes and a hexapole-like mode (Figs. \ref{fig:PWEbands}(c-e)). These are the defect modes that are predicted for the honeycomb lattice through simple symmetry arguments \cite{PainterPRBSym03}. Of the defect modes present in the bandgap, the dipole-like modes are more concentrated in the center of the defect region, thus giving them the most overlap with the QC gain material.  As such, these are the cavity modes we will focus on in the 3D simulations described below.  

Our choice of hole radius ($r/a=0.24$) in the above cavity design was made based upon a trade-off between the extent of the photonic bandgap and cavity mode localization, with that of the electrical and thermal resistance incurred in a semiconductor realization of such a structure.  As already mentioned, the reduced connectivity of a photonic lattice makes electrical injection of a defect cavity more challenging.  Electrical current in such structures is typically injected from the edge of the photonic crystal vertically through the device active region.  Injection into the defect region of the cavity is a result of lateral current spreading into the center of the photonic crystal.  The increased lateral resistance of highly porous photonic lattices significantly reduces the injection efficiency of the laser, resulting in added heating and reduced gain.  Other injection geometries are possible, such as the use of surface-plasmon laser cavities in which the top metal contact extends over the entire photonic crystal and provides not only vertical waveguiding but also electrical injection.  This geometry was used in the demonstration of the first electrically injected 2D photonic crystal microcavity laser\cite{Colombelli03}.  Nonetheless, high-aspect ratio 2D photonic crystal lattices of high porosity pose a greater challenge to fabricate due to the reduced critical dimension size.  It should be noted that further optimization of the honeycomb lattice can be obtained through use of air holes of a modified geometry.  Lattices formed from air holes with a truncated circular cross-section were found to have larger photonic bandgaps for a given critical dimension than that of the standard lattice.  In what follows, however, we focus on the honeycomb lattice with circular air holes. 

\section{3D-FDTD analysis of mid-IR devices}
\label{sec:3D-mir}

The PWE method allows the rapid solution of structure eigenmodes in the frequency domain.  However, for the 2D analysis
described above one can only approximate the laser active region with an effective index of refraction. The objective of
this section is to verify the design developed with the 2D model in Section \ref{sec:2D}, and to apply it to a realistic
mid-IR QC laser structure. An accurate representation of the structure waveguide in the epitaxial direction will be
taken into account within a 3D finite-difference time-domain (FDTD) approach. It is well known that the material system
of choice for mid-IR QC lasers is InGaAs/AlInAs lattice matched to InP \cite{qc-mid-ir1,qc-mid-ir2}. In addition,
waveguides based on surface-plasmons \cite{bahriz_apl06} have been shown to be advantageous for PC QC lasers in the
mid-IR \cite{iop04}. This is therefore the model system that we will use for the 3D simulations.

\subsection{Mid-IR surface-plasmon waveguides for QC lasers}
Semiconductor diode lasers and conventional QC lasers rely on optical waveguides where a higher-index-core is sandwiched between thick cladding layers of lower refractive index, thus confining the light inside the active region stack. The small refractive index difference between the active core and the waveguide claddings results in a standard mid-IR QC laser which is typically $6$ to $9$ $\mu$m thick. 

\begin{figure}[htb]
\centering
\includegraphics[width=0.7\linewidth]{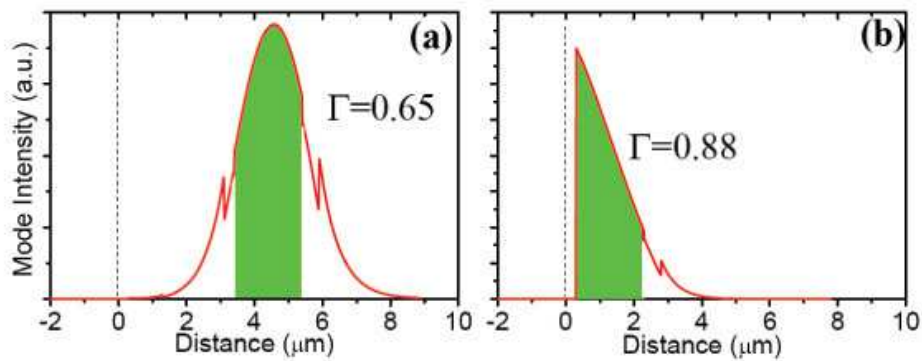}
\caption{Normalized intensity profile of the fundamental optical mode computed for a quantum cascade laser dielectric waveguide (a) and a surface-plasmon waveguide (b) designed for a wavelength of 8 $\mu$m. The shaded green areas indicate the stack of active regions and injectors. The vertical dashed line indicates the position of the device surface. In the dielectric case the thickness of the epitaxially grown material is $\approx$6 $\mu$m. The origin of the abscissa is at the air-semiconductor interface. In the surface-plasmon case the thickness is instead $\approx 3$ $\mu$m. The optical confinement factor is indicated by $\Gamma$.}
\label{fig:plasmonguide}
\end{figure}

Maxwell's equations, however, also allow for another type of optical waveguiding based upon
surface-plasmons\cite{raether}.  TM polarized electromagnetic guided modes exist at the interface of two dielectrics
with opposite sign of the real part of their dielectric constants.  Negative dielectric constants are typical of metals
below the plasma frequency. Thus, guided surface-plasmon modes at a metal-semiconductor interface are a useful
waveguiding solution for QC lasers \cite{sirtori98}. This is due to the intrinsic TM polarization, i.e. normal to the
layers, of intersubband transitions. SP waveguides need a smaller thickness of grown material, while yielding even
larger optical confinement factors $\Gamma$ (see Fig.  \ref{fig:plasmonguide}).

The surface-plasmon damping along the propagation direction can be approximated with the following formula \cite{yeh-book}:

\begin{equation}
\label{eq:SP_loss_coeff}
\alpha \approx \frac{4 \pi}{\lambda} \cdot \frac{n_{m}\cdot n_{d}^{3}}{k_{m}^{3}} ,
\end{equation}

\noindent where $k_{m}$ ($n_{m}$) is the imaginary (real) part of the metal index of refraction, $n_{d}$ is the real part of the semiconductor index of refraction, and $\lambda$ is the wavelength.  The $1/k_{m}^{3} \lambda$ dependence of the
propagation losses shows that SP waveguides are especially appealing at long wavelengths. In particular the
$1/k_{m}^{3}$ factor, which pushes the field out of the metal region and reduces the losses, becomes very small at long
wavelengths. A simple Drude model shows that the imaginary part of the index of refraction, for a generic metal,
increases dramatically when moving from short ($\lambda = 1$-$3$ $\mu$m) to long ($\lambda= 100$-$200$ $\mu$m)
wavelengths. Recent advances have also shown that low-loss metallic waveguides can be implemented at mid-IR wavelengths
\cite{bahriz_apl06}.  We therefore employ this waveguide structure as a model system for the 3D numerical simulations of
our QC laser structures. The corresponding layer sequence is displayed in Table \ref{tab:struct_ir}, together with the
corresponding indeces of refraction that are used for the simulations.

\renewcommand{\arraystretch}{1.2}
\renewcommand{\extrarowheight}{0pt}
\begin{table}
\caption{Layer structure for the mid-IR SP QC laser\cite{bahriz_apl06} modeled in the 3D FDTD simulations.  Nominal operating wavelength is $\lambda\approx 8$ $\mu$m.}
\label{tab:struct_ir}
\begin{tabularx}{\linewidth}{YYYYY}
\hline
\hline
Material & Doping (cm$^{-3}$) & Thickness ($\mu$m)  & $n$ & Function \\
\hline 
air     & n.a. & 4 & 1 & top cladding \\
Au     & n.a. & 0.3 & perfect conductor & metal contact \\
  InGaAs & 10$^{17}$/10$^{18}$ & 0.05   & 3.3/3.47  & contact layers     \\ 
  InGaAs-AlInAs MQW structure & 1.7 x 10$^{16}$  & 2.6 & 3.374  & active region  \\
  InGaAs & 5 x 10$^{16}$ & 0.5 & 3.475 & Buffer/lower cladding layer \\
  InP    & 10$^{17}$ & 3 &  3.077 & substrate \\
\hline
\hline
\end{tabularx}
\renewcommand{\arraystretch}{1.0}
\end{table}

\begin{figure}[htb]
\centering
\includegraphics[width=0.8\linewidth]{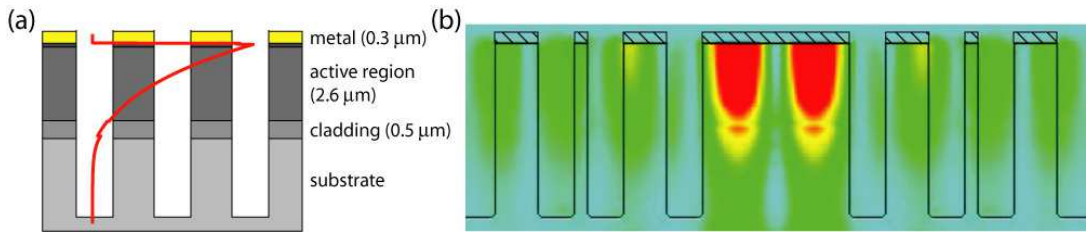}
\caption{(a) Surface-plasmon optical mode profile superimposed on a sketch of the vertical section of a mid-IR photonic crystal QC laser with air holes extending $4.7$ $\mu$m deep into the semiconductor.  (b) 3D-FDTD calculated electric field intensity cross-section of a defect mode (top-view shown in Fig. \ref{fig:FDTD-mir}(a)) of the deeply patterned honeycomb photonic lattice. The vertical mode profile shows a similar behavior as the 1D surface-plasmon simulation of (a), whereas the confinement in the in-plane direction is provided by distributed Bragg reflection of the photonic lattice.}
\label{fig:etched}
\end{figure}

While the vertical confinement is provided by the SP waveguide, the in-plane optical confinement is induced by the
photonic crystal. A high index contrast can be obtained by penetration of the air holes through the top metal layer and
deep into the semiconductor waveguide structure.  Intuitively, reduced scattering losses are obtained for a hole depth
which overlaps a significant fraction of the guided mode energy\cite{durso98}.  In Fig. \ref{fig:etched}(a) we show a
schematic of the vertical cross-section of the photonic crystal SP waveguide structure, and in Fig. \ref{fig:etched}(b) we plot the
vertical mode profile of a localized defect mode (see Fig. \ref{fig:FDTD-mir}(a)) of the honeycomb lattice of air holes
with a hole depth of $4.7$ $\mu$m in the semiconductor heterostructure.  At an operating wavelength of $8$ $\mu$m this
results in a vertical mode overlap (energy overlap) with the air holes of almost $90\%$.  It is clear from this example
why surface-plasmon QC lasers are ideally suited to PC technology; their reduced thickness allows for a significantly
shallower etch of the semiconductor material in comparison to conventional laser waveguide structures in which an etch
depth of $10$ $\mu$m or more would be required in the mid-IR.

\subsection{Defect cavity design}
The cavity investigated in this section is the same as in Section \ref{sec:supercell}.  It is obtained by removing a full
hexagon of air holes from the PC lattice.  As depicted in Fig. \ref{fig:etched}, we ``etch'' 4.7-$\mu$m deep holes into the
semiconductor laser structure, with the $r/a$ ratio of the air holes set to $0.21$.  The properties of the different
laser structure layers used in the simulation are given in Table \ref{tab:struct_ir}.  Full 3D-FDTD
simulations\cite{PainterPRBSym03,PainterJOSAB99,durso98} were performed with an effective grid resolution of 100 nm,
corresponding to roughly 20 points per wavelength in the high-index semiconductor material at an operating wavelength of
$8$ $\mu$m.  The use of perfectly conducting boundaries for the top metal contact layer is a small approximation at $8$ $\mu$m due to the much higher plasma frequency of the Au metal contacts used in practice.  

\begin{figure}[htb]
\centering
\includegraphics[width=0.8\linewidth]{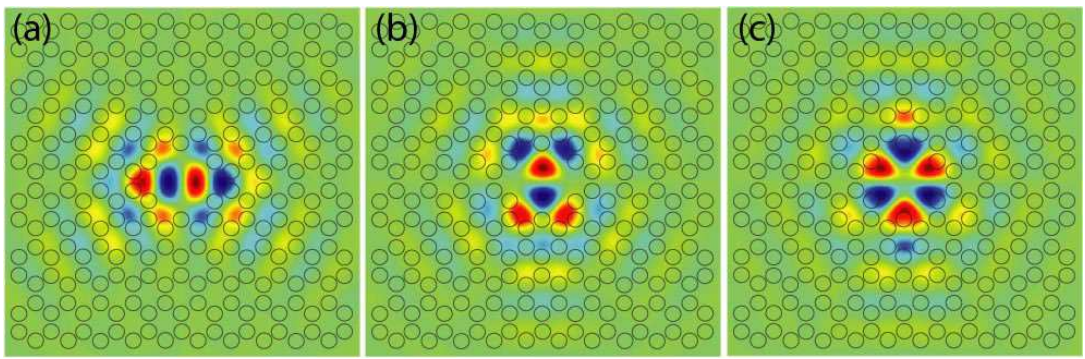}
\caption{In-plane mode profiles ($E_{z}$) for the (a) $x$-dipole, (b) $y$-dipole, and (c) hexapole defect modes of the surface-plasmon vertical waveguide structure with an ``etched'' hexagonal defect cavity in the honeycomb lattice.}
\label{fig:FDTD-mir}
\end{figure}

A simulation of a photonic crystal cavity with $n_{x}=7$ periods in the $\hat{x}$-direction and $n_{y}=4$ periods in the
$\hat{y}$-direction yields the three localized defect modes shown in Fig.  \ref{fig:FDTD-mir}(a-c).  These modes are the
same modes as found in the 2D analysis of Fig. \ref{fig:PWEbands}.  Comparison of the 2D and 3D mode properties are
summarized in Table \ref{tab:defect-energies}.  The normalized frequencies of the defect modes lie within the TM-like
bandgap of the 3D honeycomb lattice waveguide structure, and are centered around $a/\lambda \sim 0.2$.  At a wavelength
of $8$ $\mu$m this corresponds to a lattice constant of approximately $a=1.6$ $\mu$m, a hole radius of $r=0.34$ $\mu$m,
and a minimum dimension given by the gap between nearest neighbor holes of $a(1/\sqrt{3} -2r/a)=0.25$ $\mu$m.

Effective quality factors associated with in-plane ($Q_{\parallel}$), topside vertical ($Q_{t}$), and bottomside
vertical ($Q_{b}$) radiation losses are also calculated for the 3D FDTD simulations.  The topside vertical effective
$Q$-factor for all of the defect modes is estimated at $> 10^{6}$, due to the fact that the cavity modes lie
predominantly below the light cone of the top air-cladding\cite{ref:Srinivasan1}.  $Q_{b}$ on the other hand, can be
increased only slightly from that reported in Table \ref{tab:defect-energies} through deeper etching of the air holes,
limited mainly by the high-index of the semiconductor substrate and the localized nature of the cavity.  Due to the
TM-like polarization in-plane bandgap provided by the honeycomb lattice, an increase in the number of PC mirror periods
to a value above $10$ effectively eliminates in-plane radiation loss, and the $Q$-factor of the cavity modes is limited
by both radiation into the substrate as well as material absorption loss due to ohmic heating in the metal
surface-plasmon layer (for wavelengths in the mid-IR, and a Au metal surface-plasmon layer, material absorption limits
the $Q$-factor to approximately $10^3$\cite{kartik_apl04}).

\renewcommand{\arraystretch}{1.2}
\renewcommand{\extrarowheight}{0pt}
\begin{table}
\begin{center}
\caption{Comparison between 2D (PWE) and 3D (FDTD) simulated defect modes.}
\label{tab:defect-energies}
\begin{tabularx}{\linewidth}{YYYY}
\hline
\hline
Parameter & $x$-dipole mode & $y$-dipole mode & hexapole mode \\
\hline 
PWE $a/ \lambda$ & 0.2115 & 0.2115 & 0.2162 \\
FDTD $a/ \lambda$ & 0.196 & 0.197 & 0.205 \\
$Q_{\parallel}$ & 326 & 288 & 169 \\  
$Q_{t}$ & $1.2\times 10^7$ & $9\times 10^6$ & $1.1\times 10^7$  \\  
$Q_{b}$ & 92 & 93 & 114  \\  
\hline
\hline
\end{tabularx}
\renewcommand{\arraystretch}{1.0}
\end{center}
\end{table}


\section{Analysis of THz devices}
\label{sec:3D-thz}

Our FDTD analysis of mid-IR PC QC lasers examined structures in which the honeycomb lattice was patterned through the QC semiconductor active region. This is necessary in order to induce a strong index contrast. However, a different solution is possible if the optical mode is highly confined in the vertical direction, as is the case for THz QC lasers with metal-metal waveguide structures\cite{unterrainer02,williams_apl03}. In this case, and under certain constraints related to the structure thickness, the sole patterning of the top metal contact is capable of inducing a complete photonic bandgap.  These metal-insulator-metal (MIM) structures\cite{Prade91}, as we will refer to them here (see Figs. \ref{fig:double-metal} and \ref{fig:structschem}), will be the focus of our time domain modeling in the following sections.  The appeal of this design lies in the fact that only the top metal layer needs to be patterned\cite{ref:Mahler,ref:Demichel}, which both simplifies the fabrication process and improves the efficiency of carrier diffusion in the active region.

\subsection{MIM structures: waveguides for THz QC lasers}
\label{subsec:MIM}
\begin{figure}[htb]
\centering
\includegraphics[width=0.8\linewidth]{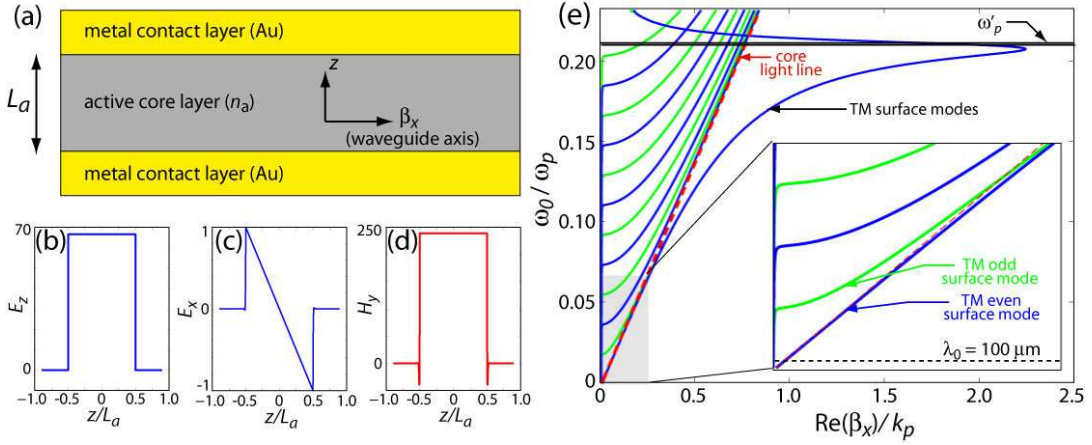}
\caption{(a) Schematic showing the cross-section of a MIM waveguide (metal layers are assumed semi-infinite).  (b) $E_z$ normal electric field component, (c) $E_x$ longitudinal field component, and (c) $H_{y}$ magnetic field plot of the even parity guided surface mode at $\lambda=100$ $\mu$m for an active region core thickness of $L_{a}=1$ $\mu$m.  (e) Dispersion diagram of the double-metal (Au) waveguide structure ($L_{a}=1$ $\mu$m).  The core is modeled with a constant refractive index of $n_{a}=3.59$, while the metal layers are modeled using a Drude-Lorentz model for Au (background dielectric constant $\epsilon_{b}=9.54$, plasmon frequency $\omega_{p}=1.35 \times 10^{16}$ rad/s, and relaxation time $8 \times 10^{-15}$ s).}
\label{fig:double-metal}
\end{figure}

Metal-metal waveguides\cite{Prade91} are usually employed for QC lasers in the THz range\cite{unterrainer02,williams_apl03} because they can provide almost unity optical confinement factors, and simultaneously, relatively low waveguide losses. The active laser core is sandwiched between two metal layers (typically Ti/Au or Ge/Au/Ni/Au) which act as surface-plasmon layers (Fig. \ref{fig:double-metal}(a)).  The two surface-plasmon modes of both the top and bottom metal-core interfaces become coupled and form two guided modes, one of even parity and one of odd parity (assuming a nearly symmetric metal-core-metal structure, and where the \emph{vector} parity is determined by the symmetry of the magnetic field).  The dispersion diagram of a double-metal waveguide with active layer core thickness of $L_{a}=1$ $\mu$m is displayed in Fig. \ref{fig:double-metal}(e).  For sub-wavelength core thicknesses the odd parity surface mode is cut-off, while the even parity mode exists all the way down to zero frequency.  The electric and magnetic field profiles of the even parity surface mode at a wavelength of $\lambda_{0}=100$ $\mu$m are shown in Fig. \ref{fig:double-metal}(b-d), showing the near-unity confinement factor of the fields in the core region.      

As has been noted by other authors\cite{MaierOpEx06}, a reduction in the dielectric core region thickness of MIM structures results in an increase in the in-plane wavevector of the even parity guided mode for a particular frequency.  This is a result of the ``pushing'' of the electromagnetic energy into the metal cladding regions with decreased core thickness.  Of import to the current work is the ability to create a large effective index contrast through patterning only of the metal layers.  In such a patterned structure the guided mode sees two effective indeces as it propagates, the large effective index of the guided surface-plasmon mode where the metal is left intact, and a lower effective index, more delocalized mode of the dielectric core and air-cladding where the metal is removed. That this picture is in fact valid, and can be used to great effect, is demonstrated below where we analyze the bandstructure and localized resonances of double-metal waveguides with only a surface-patterning in the top metal contact.              

\subsection{Patterned MIM waveguide bandstructure analysis}
\label{subsec:THz_bandstructure}

\begin{figure}[ht]
\centering
\includegraphics[width=0.7\linewidth]{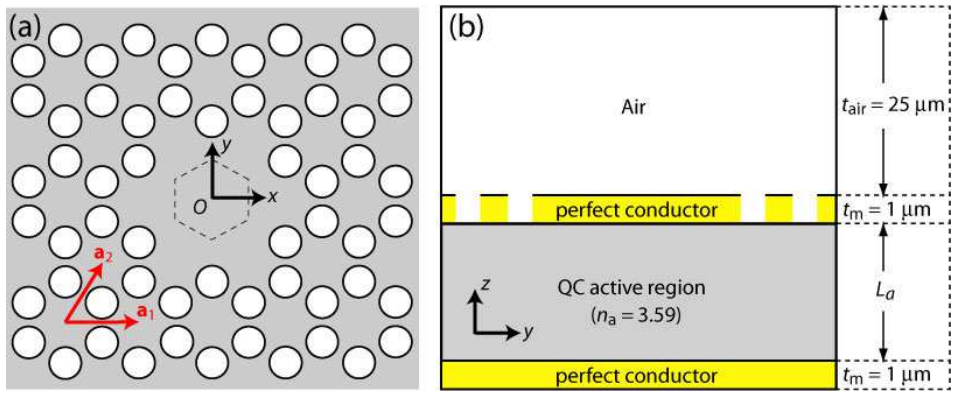}
\caption{Schematic of the honeycomb lattice and MIM structure used in FDTD modeling. (a) Top view of the lattice with a central defect consisting of the removing of the central hexagon of air holes. (b) Cross-section of the simulated structure. Only the top metal layer (perfect conductor) is patterned.  The displayed layer thicknesses are for an operating wavelength of $\lambda_{0}=100$ $\mu$m and a normalized frequency of $a/\lambda_{0}=0.17$ within the bandgap of the honeycomb lattice.}
\label{fig:structschem}
\end{figure}

The model MIM structure that we will consider here, shown schematically in Fig. \ref{fig:structschem}(b), consists of the following sequence of layers (from bottom to top):  a bottom unpatterned metal contact, a QC active region dielectric core (modeled with a uniform index of $n_{a} = 3.59$ for simplicity\cite{Kohen05}), a top patterned metal contact, and a region of air above the structure.  The metal layers are assumed to be perfect conductors; loss due to absorption in the metal regions is discussed in section \ref{subsec:THz_losses}.  Mur absorbing boundary conditions\cite{ref:Mur} are used to model radiation loss out of the structure.  For each of the THz FDTD simulations described below, a grid resolution of $58$ points per lattice constant $a$ of the honeycomb lattice was used.  At a nominal operating wavelength of $100$ $\mu$m, and for normalized frequencies $a/\lambda_{0} \sim 0.17$ within the bandgap of the honeycomb lattice (see below), the lattice constant is $a=17$ $\mu$m.   At this wavelength and normalized frequency (hereinafter the \emph{nominal operating conditions}), the corresponding spatial resolution is $0.3$ $\mu$m (roughly $95$ points per wavelength in the dielectric core material), and the MIM layer thicknesses are as shown in Fig. \ref{fig:structschem}(b).   

\begin{figure}[htb]
\centering
\includegraphics[width=0.5\linewidth]{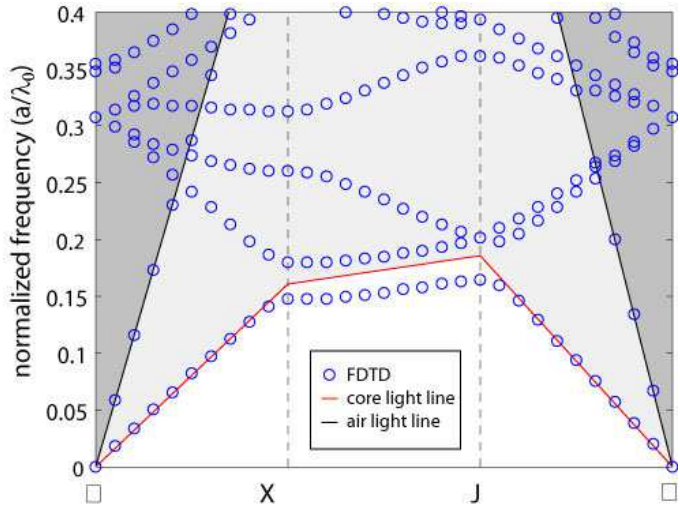}
\caption{FDTD calculated bandstructure of a patterned MIM waveguide with a normalized dielectric core thickness of $\overline{L}_{a}=L_{a}/a=0.176$ and a normalized air hole radius of $r/a=0.25$ in the top metal contact.  Metal layers are simulated as perfect conducting boundaries.  At an operating wavelength of $\lambda_{0}=100$ $\mu$m within the photonic bandgap ($a/\lambda_{0}=0.17$), the corresponding physical sizes of the patterned double-metal waveguide are $\{a,\ r,\ L_{a}\}=\{17,\ 4.25,\ 3\}\mu$m.} 
\label{fig:3umbandstruct}
\end{figure}

The band diagram for surface-patterned MIM waveguide structures with various dielectric core thicknesses were computed to confirm the existence of a complete in-plane bandgap, and to examine the dependence of the bandgap frequency width on structure thickness.  The simulation volume consists of a single unit cell of the honeycomb lattice with Bloch periodic boundary conditions applied in the plane of periodicity and Mur boundary conditions in the top and bottom directions (normal to the metal and semiconductor layers).  Figure \ref{fig:3umbandstruct} shows the band diagram for a structure with a normalized core thickness of $\overline{L}_{a}=L_{a}/a=0.176$ and a normalized air hole radius of $r/a=0.25$ in the top metal layer.  The corresponding physical core thickness is $L_{a}=3$ $\mu$m  under nominal operating conditions ($\lambda_{0}=100$ $\mu$m, $a/\lambda_{0}=0.17$).  In the band diagram, truly guided modes lie below the \emph{air} light-line, while leaky modes lie above it (in this case light can only leak into the top vertical direction where patterning of the metal has been applied).  The bottom-most band is a surface wave, lying just below the light-line of the active region core.  It is akin to the even parity guided surface-plasmon mode of the symmetric (unpatterned) double-metal waveguide studied above.  The higher-lying frequency bands constitute zone-folded versions of this surface wave.  An in-plane photonic bandgap exists for guided modes of this structure between normalized frequencies $a/\lambda_{0}=0.165-0.18$.  Due to the extreme thinness of the dielectric core, the higher-order vertical modes of the MIM waveguide are not present in this diagram, but lie at much higher frequencies ($a/\lambda_{0} > 0.75$).  As a result, the photonic bandgap is a true full in-plane bandgap for guided resonances of the MIM structure, not just for a single polarization or mode symmetry.  Note, the modes that lie along the air light-line in the diagram are radiation modes, predominantly localized in the air regions above the MIM structure.     

\begin{figure}[ht]
\centering
\includegraphics[width=0.5\linewidth]{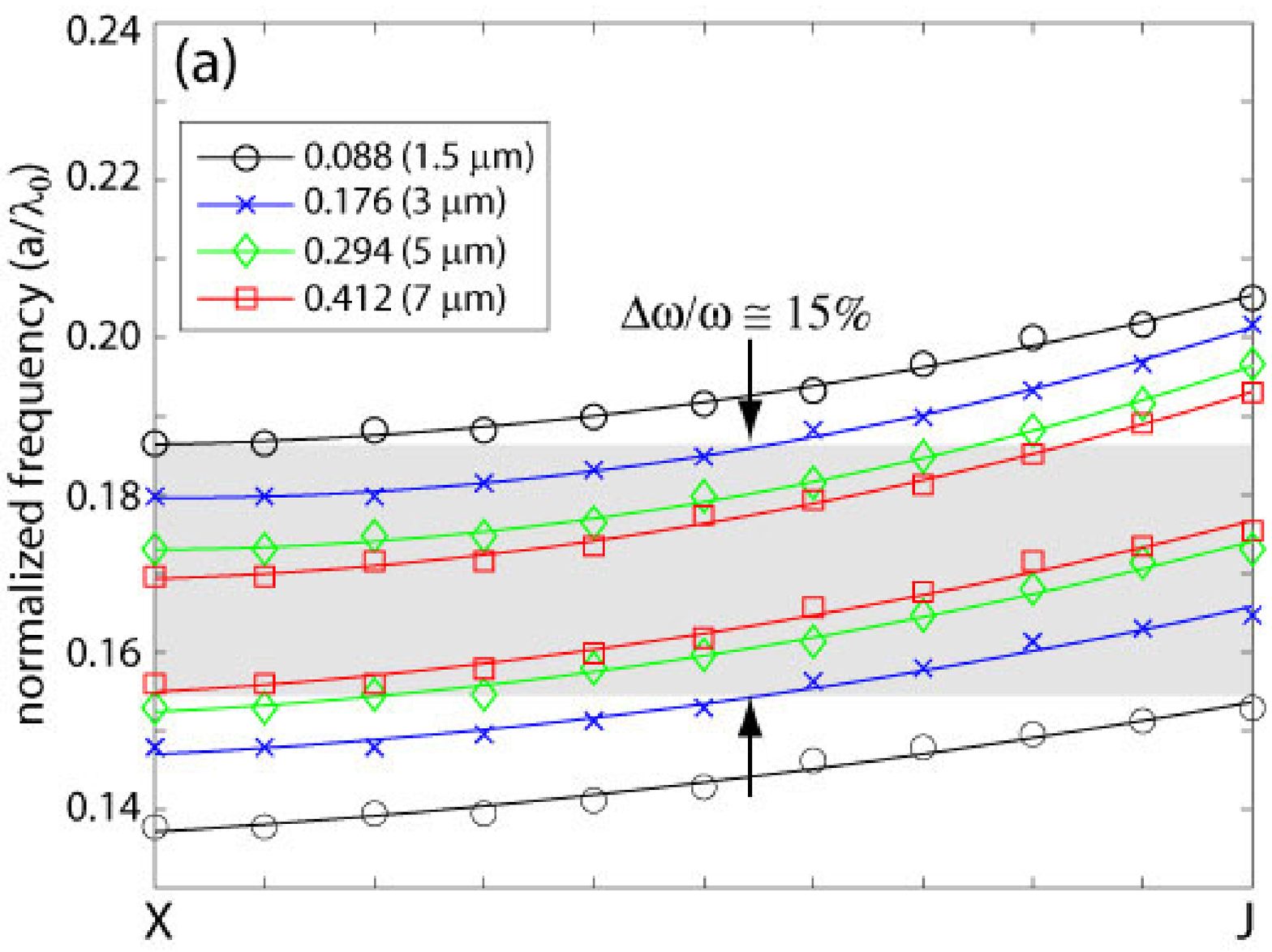}
\caption{(a)  Valence and conduction band dispersion between the $X$ and $J$ points for varying active region core thickness ($\overline{L}_{a}=\{0.088 (\textcolor{black}{\circ}),0.176 (\textcolor{blue}{\times}),0.294 (\textcolor{green}{\diamond}),0.412 (\textcolor{red}{\square})\}$; nominal physical thicknesses $L_{a}=\{1.5 (\textcolor{black}{\circ}),3 (\textcolor{blue}{\times}),5 (\textcolor{green}{\diamond}),7 (\textcolor{red}{\square})\}$ $\mu$m).  The symbols correspond to data points from FDTD simulations and the lines are guides to the eye.  The normalized hole radius for all simulations was fixed at $r/a=0.25$.  Note that the complete band gap shrinks as device thickness increases, and is closed for $\overline{L}_a>0.294$ ($L_a>5$ $\mu$m).} 
\label{fig:bandgaps}
\end{figure}

\begin{figure}[ht]
\centering
\includegraphics[width=0.7\linewidth]{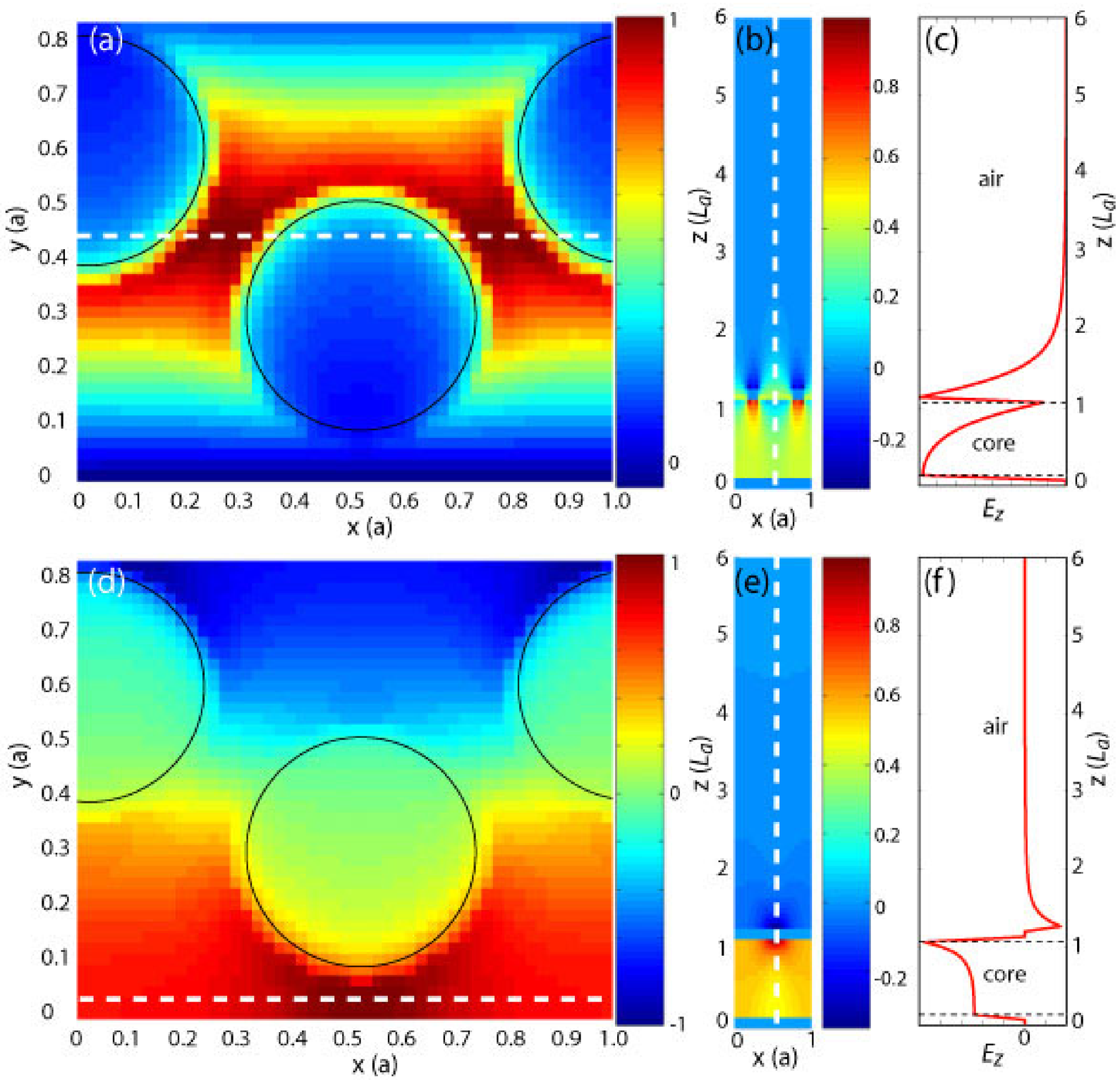}
\caption{High-frequency (a-c) and low-frequency (d-f) gap mode field plots ($E_{z}$) at the $X$-point of the honeycomb lattice for the patterned double-metal waveguide structure of Fig. \ref{fig:bandgaps} with $\overline{L}_{a}=0.294$ ($L_{a}=5$ $\mu$m).  The in-plane field plots of (a,d) correspond to a section through the core of the MIM structure just below the patterned top metal contact.  The dashed white lines in (a,d) and (b,e) indicate the position of the sections used in the field plots of (b,e) and (c,f), respectively.} 
\label{fig:X_point_gapmodes}
\end{figure}

In Fig. \ref{fig:bandgaps} we plot the dispersion of the bands defining the (lowest) photonic bandgap between the high symmetry $X$ and $J$ points of the honeycomb lattice for varying dielectric core thicknesses.  The bandgap is seen to shrink with increasing dielectric core thickness, closing for a normalized active region thickness greater than $\overline{L}_{a}=0.294$ (nominal physical thickness $L_{a}=5\ \mu$m).  Mode field plots (see Fig. \ref{fig:X_point_gapmodes}) at the $X$ and $J$ points indicate that the high and low frequency modes defining the bandgap are of mixed type: the low-frequency ``dielectric'' or ``valence'' band mode is predominantly a surface wave at the interface of the top patterned metal surface, whereas the high-frequency ``air'' or ''conduction'' band mode is predominantly a surface wave of the bottom unpatterned metal interface.  Additionally, the high-frequency gap mode sits largely beneath the unpatterned regions of the top metal boundary and the low-frequency mode resides below the patterned areas with more energy residing in the air cladding.  The photonic bandgap in this case is a result of differences in these two surface waves.  This simple picture, although correct, betrays its complexity for two reasons: (i) perfect metal conductor boundaries were used in the simulations, which do not support surface waves when continuous and flat, and (ii) the patterned air holes are much smaller than the wavelength in air of the guided mode.  That a surface wave can indeed exist at the interface of a \emph{patterned} perfect conducting layer was pointed out in Ref. \cite{Pendry04}, where it was shown that it is precisely the cut-off frequency of the air holes in the perfect conductor which sets the effective plasma frequency of the layer.  The thinner the dielectric core layer, the larger the fraction of energy that resides at the top patterned metal interface, and a greater degree of mode delocalization that occurs into the air-cladding where the top contact is removed, both effects which increase the effective index contrast of the two surface waves and thus the photonic bandgap.  It should be noted that the use in our model of perfect conducting boundaries as opposed to a Au metal contact layer, for instance, does not affect the photonic bandstructure shown in Fig. \ref{fig:3umbandstruct} and Fig. \ref{fig:bandgaps}, as the plasma frequency of Au is much greater than the THz frequencies under consideration.  Further discussion and analysis of photonic bandgap structures formed from surface patterned metal and perfectly conducting layers will be presented in a future study\cite{ref:Painter_spoof_bandgap}.  In practice, laser structures with thicknesses much below that of several microns may be difficult to fabricate and incur insurmountable optical losses (see below for discussion), but the trend shown in Fig. \ref{fig:bandgaps} clearly indicates that such structures would support complete bandgaps that are quite substantial, approaching $15\%$ of the gap center frequency for $L_{a}=1.5$ $\mu$m at an operating wavelength of $\lambda_{0}=100$ $\mu$m.

\subsection{Characterization of defect modes}
\label{subsec:THz_defect_modes}
\begin{figure}[htb]
\centering
\includegraphics[width=0.6\linewidth]{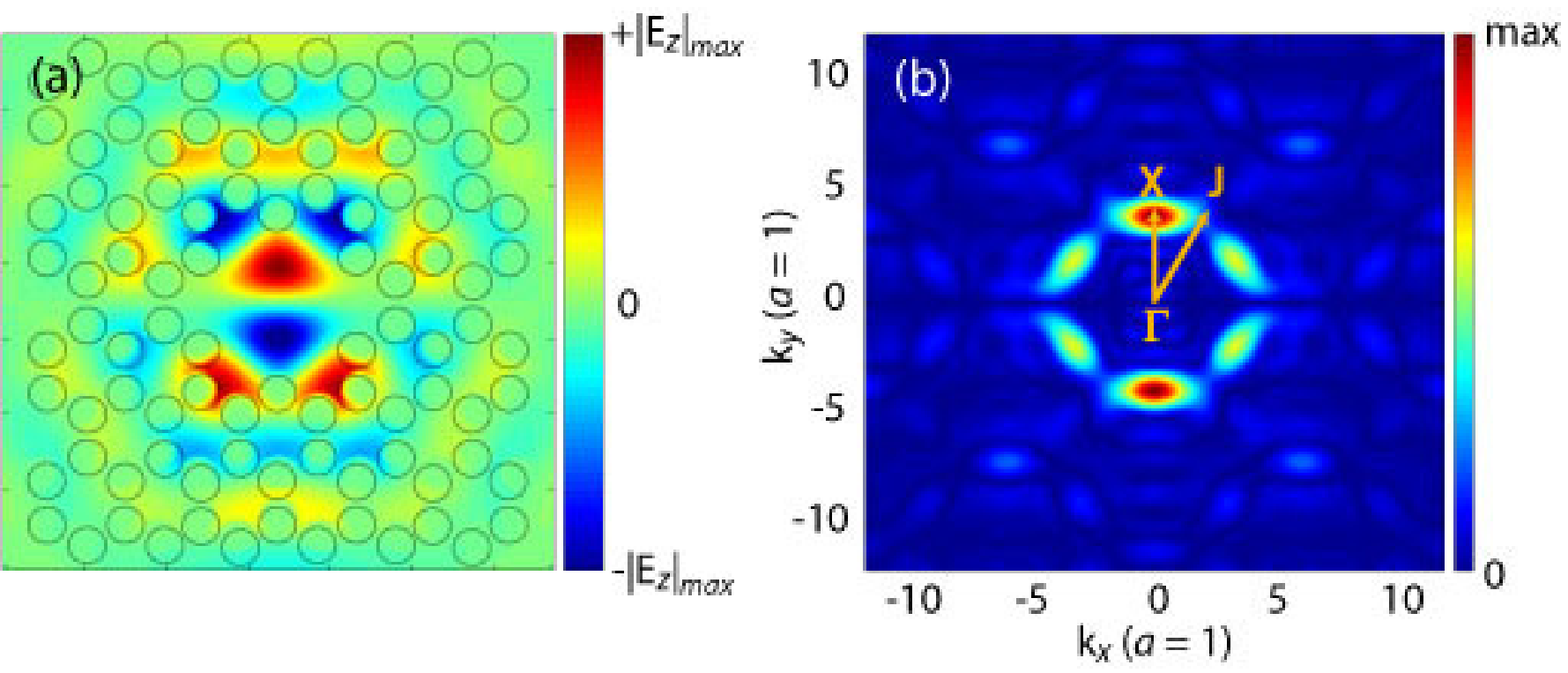}
\caption{In-plane mode profile of the $\hat{y}$-polarized dipole-like defect mode for a point defect consisting of the removal of the central 6 air-hole patterns in the top metal contact (see Fig. \ref{fig:structschem}).  (a) Real space profile of $E_{z}$ and (b) magnitude of the spatial Fourier transform of $E_{z}$ for an in-plane cut through the middle of the MIM waveguide.} 
\label{fig:dipolemode}
\end{figure}

Ultimately, one would like to utilize the strong photonic bandgap effects present in the patterned MIM waveguide structures studied above to, for instance, form localized microcavity laser resonators.  In order to demonstrate the effectiveness of the top metal patterning in this regard, we have also simulated the localized resonances that are formed within and around point defects of the honeycomb metal pattern studied in the previous sub-section.  In Fig. \ref{fig:dipolemode}(a) we show the $E_{z}$ field plot of the localized $\hat{y}$-dipole-like\footnote{The symmetry of the mode is that of a $\hat{y}$-polarized dipole in the plane of the MIM waveguide.  It is degenerate with a second dipole-like mode with $\hat{x}$-polarization\cite{PainterPRBSym03}.} mode that forms around a point-defect consisting of the removal of the central 6 holes in the top metal (see Fig. \ref{fig:structschem}(a)) of a MIM waveguide with an active region thickness of $L_{a}=3$ $\mu$m ($\overline{L}_{a}=0.176$).  The corresponding (in-plane) spatial Fourier transform of the mode is plotted in Fig. \ref{fig:dipolemode}(b), showing that the mode is localized at the $\vecb{k}_{X}$ points in reciprocal space, and consistent with a ``donor'' mode formed from the ``conduction'' bandedge at the $X$-point\cite{PainterPRBSym03}.  This mode is identified as the same dipole-like mode of the mid-IR surface-plasmon simulations of Fig. \ref{fig:FDTD-mir}(b) in section \ref{sec:3D-mir}, where in those structures a deep air hole patterning of the semiconductor dielectric material was used to create a large index-contrast.    

\begin{figure}[htb]
\centering
\includegraphics[width=0.5\linewidth]{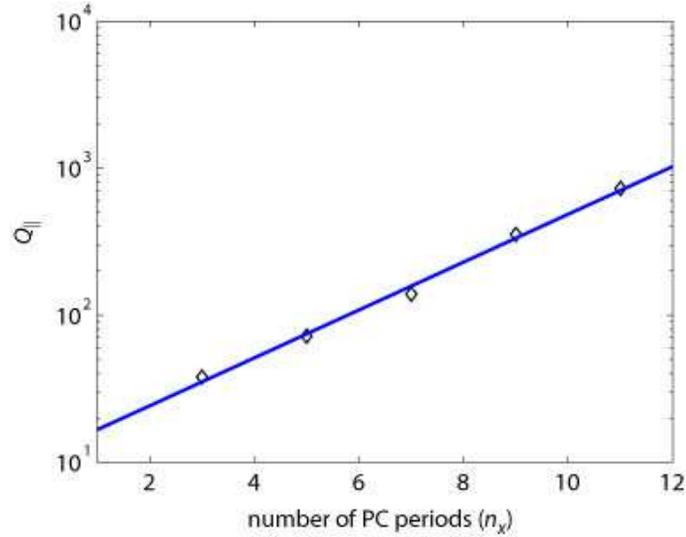}
\caption{Increase of $Q_{\parallel}$ with the number of PC periods for an MIM structure with $L_{a}=3$ $\mu$m ($\overline{L}_{a}=0.176$) thick active region.  The diamonds show the results of FDTD simulations and the solid line is a least squares fit to these results.  The PC cavities were chosen to be approximately square, so that the number of lattice periods ($n_x \times n_y$) used in the FDTD simulations were $3\times2$, $5\times3$, $7\times4$, $9\times5$ and $11\times6$ (referring to Fig. \ref{fig:structschem}(a), a ``period'' in the $\hat{x}$ direction is taken as $a$, whereas in the $\hat{y}$ direction it is $\sqrt{3}a$).  The abscissa of the plot refers to the number of periods ($n_{x}$) in the $\hat{x}$ direction.}
\label{fig:Qparvsperiods}
\end{figure}

For the fixed waveguide core thickness of $3$ $\mu$m the $\hat{y}$-dipole-like mode of Fig. \ref{fig:dipolemode}(a) is seen to be highly localized in real-space, as expected from the bandgap simulations of Fig. \ref{fig:bandgaps}.  Additional confirmation of the in-plane localization of the defect mode due to an in-plane photonic bandgap effect can be obtained by studying the scaling of the in-plane radiation losses as a function of the number of periods of the PC patterning surrounding the central defect.  The total quality factor due to radiation loss is determined by 

\begin{equation}
\frac{1}{Q_{\text{rad}}} = \frac{1}{Q_{\parallel}} + \frac{1}{Q_{\perp}},
\end{equation}

\noindent where $Q_{\parallel}$ and $Q_{\perp}$ quantify the in-plane and out-of-plane radiation losses\cite{PainterJOSAB99}. The out-of-plane quality factor is found to be $> 10^5$ for these structures; thus the dominant radiation loss occurs in the in-plane direction.  Fig. \ref{fig:Qparvsperiods} shows a plot of the $Q$-factor as a function of the number of surrounding PC periods.  The radiation $Q$-factor increases exponentially with period number as expected for a bandgap mode, with a cavity consisting of only $12\times6.5$ periods able to sustain a mode with radiation $Q > 10^3$.


\subsection{Effective mode volume and metal absorption losses}
\label{subsec:THz_losses}

We have shown above that in double metal structures the frequency-width of the photonic bandgap of an honeycomb photonic lattice {\it increases} when the structure is thinner. It implies, intuitively, that the photonic crystal becomes increasingly effective in confining light when the waveguide core thickness is reduced. To quantify the modal localization versus core region thickness we calculated the effective mode volumes $V_{\text{eff}}$ of the $\hat{y}$-dipole-like mode of Fig. \ref{fig:dipolemode} for various active region thicknesses ($L_{a}$). $V_{\text{eff}}$ is defined as:
\begin{equation}
V_{\text{eff}}=
\frac{\int \epsilon |{\bf E}|^2 dV}{max\left[\epsilon |{\bf E}|^2 \right]},
\end{equation}
\noindent where $max[..]$ denotes the maximum value of the argument. The results are reported in Table \ref{tab:Veff}.  The decrease of $V_{\text{eff}}$ with the active core region thickness $L_{a}$ is super-linear, as can be seen in the third column of Table \ref{tab:Veff} in which $V_{\text{eff}}/L_{a}$ is tabulated.  The super-linear decrease in $V_{\text{eff}}$ with active region thinning is a clear indication that the photonic crystal formed from patterning of the top metal surface provides more effective in-plane confinement for thinner waveguide cores. 

\renewcommand{\arraystretch}{1.2}
\renewcommand{\extrarowheight}{0pt}
\begin{table}
\begin{center}
\caption{Effective volume ($V_{\text{eff}}$), metal-absorption-limited loss coefficient ($\alpha_{m}$), $Q$-factor ($Q_{m}$), and guided wave energy velocity $\nu_{\text{E}}$ as a function of the active region thickness at a nominal vacuum wavelength of $\lambda=100\ \mu$m.}
\label{tab:Veff}
\begin{tabularx}{\linewidth}{YYYYYY}
\hline
\hline
$L_{a}$ ($\mu$m) & $V_{\text{eff}}$ $\bigl((\lambda_{0}/n_{a})^3\bigr)$ & $V_{\text{eff}}/L_{a}$ $\bigl((\lambda_{0}/n_{a})^2\bigr)$& $\alpha_{m}$ $\bigl($cm$^{-1}\bigr)$ & $c/\nu_{\text{E}}$ & $Q_{m}$  \\ 
\hline 
1.5 & 0.37 & 0.247 & 50.6 & 3.627 & 45 \\
3 & 0.93 & 0.31 & 25.8 & 3.609 & 88 \\
3.5 & 1.18 & 0.337  & 22.2 & 3.606 & 102 \\
5 & - & - & 15.5 & 3.601 & 146 \\  
7 & - & - & 10.5 & 3.598 & 215 \\  
\hline
\hline
\end{tabularx}
\renewcommand{\arraystretch}{1.0}
\end{center}
\end{table}

FDTD simulations to this point have approximated the metal contacts defining the MIM waveguide structures with perfect conductor boundary conditions, neglecting ohmic heating loss in the metal layers. Additionally, the optical losses within conventional THz QCLs include a contribution from free-carrier absorption, the level of which can be minimized through reduction of the thickness and doping levels of the semiconductor contact layers\cite{unterrainer02,williams_apl03,Kohen05}.  Given that the thin MIM waveguide structures studied here have a large fraction of mode energy at the metal surfaces, we will concern ourselves here with estimating the optical losses in the metal layers, with free-carrier absorption and other losses contributing at a level similar to that in conventional THz QCL structures.  Incorporation of ``real'' metals into the 3D FDTD simulations would require a prohibitively large amount of memory and/or simulation time due to the very short skin depth at THz frequencies.  Fortunately, an estimate of the ohmic heating losses in the metal can be obtained for the patterned MIM waveguide structure by considering the analytically tractable unpatterned MIM structure.  

\begin{figure}[ht]
\centering
\includegraphics[width=0.5\linewidth]{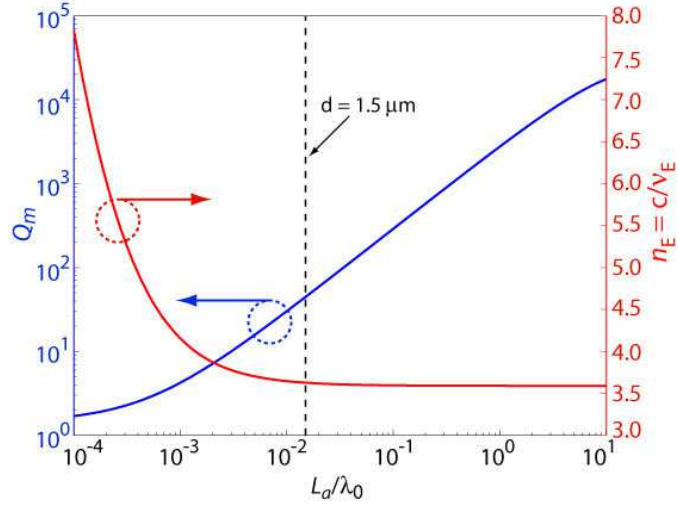}
\caption{Effective $Q$-factor due to metal waveguide loss ($Q_{m}$, blue line) and effective energy velocity index ($n_{E} \equiv c/\nu_{E}$, red line) versus active region thickness ($L_{a}$) for a 1D MIM waveguide with Au metal guiding layers at $\lambda_{0}=100$ $\mu$m.}
\label{fig:metallossQ}
\end{figure}

Specifically, we approximate the losses due to absorption in the metal contacts through an analysis of the fundamental, even parity, coupled-surface-plasmon mode in a simple planar 1D double-metal waveguide structure\cite{Prade91}. The complex dielectric constant of the Au metal layers in the THz ($\lambda = 100$ $\mu$m) are taken from Ref. \cite{Ordal83} ($\epsilon_{\text{Au}}=-1.06\times 10^5 + i 1.93\times 10^5$). The refractive index of the QC active region core (which is assumed to be homogeneous and lossless) is taken to be 3.59. The fields and complex propagation constant are determined through numerical solution of the dispersion relation.  The imaginary part of the propagation constant, which determines the modal loss per unit length down the waveguide (see Table \ref{tab:Veff}), is converted to a loss per unit time and related to an effective $Q$ due to metal absorption for comparison with the radiation losses plotted in Fig. \ref{fig:Qparvsperiods}:

\begin{equation}
Q_{m} = \frac{\omega_{0}}{2\nu_{\text{E}}\text{Im}(\beta)},
\end{equation}

\noindent where $\nu_{\text{E}}$ is the \textit{global} energy velocity of the mode in the waveguide and $\omega_{0}$ the angular frequency of the mode. The energy velocity is computed from  

\begin{equation}
\nu_{\text{E}} \equiv \frac{\int\left<S\right>\ \text{d}z}{\int\left<W\right>\ \text{d}z},
\end{equation}

\noindent where $\left<S\right>$ is the time-averaged Poynting flux, $\left<W\right>$ the time-averaged energy density valid for a lossy, dispersive medium \cite{Ruppin02}, and the quantities are integrated along the transverse width ($z$-direction) of the waveguide.  Fig. \ref{fig:metallossQ} shows the variation of  $Q_{m}$ and $c/\nu_{\text{E}}$ versus the width of the core region. This behavior is due to the penetration of the electric field into the lossy metallic layers when the active region is made thinner \cite{williams_apl03,MaierOpEx06}.  The metal-absorption-limited propagation loss per unit length ($\alpha_{m}$) and $Q$-factor ($Q_{m}$) for active core region thicknesses of $L_{a} = 1.5$, $3$, $3.5$, $5$, and $7$ $\mu$m are tabulated in Table \ref{tab:Veff} along with the effective mode volume data.  From this estimate we see that metal absorption loss will dominate the in-plane radiation losses of the $\hat{y}$-dipole-like mode of Fig. \ref{fig:dipolemode} for defect cavities of $9 \times 5$ periods or more.  

In order to connect this work with that of more conventional THz QC laser structures and measurements, and to determine whether such thin double-metal waveguide structures are at all practical, one must relate the calculated $Q$-factors of Fig. \ref{fig:Qparvsperiods} and Fig. \ref{fig:metallossQ} to per unit length exponential loss coefficients.  The typical gains that can be achieved in THz QC lasers at $\lambda\approx 100$ $\mu$m are $\sim 20-40\ \text{cm}^{-1}$ \cite{Kohen05}. These values refer to modes whose energy velocity is approximately equal to the phase velocity of the bulk material.  In thin metal waveguide structures the energy velocity of the plasmon mode can be much reduced relative to that for a mode in a thick waveguide structure (see the dispersion of Fig. \ref{fig:double-metal}(e)); however, for the MIM waveguide thicknesses considered here ($L_{a} \ge 1.5$ $\mu$m) the energy and core phase velocity are nearly equal and the metal absorption loss coefficients, $\alpha_{m}$, of Table \ref{tab:Veff} can be directly compared to measured gain coefficients of conventional laser structures.  Such comparisons indicate that for all but the thinnest waveguide structure ($L_{a}=1.5$ $\mu$m), laser action should be achievable. 

\section{Conclusions}
In conclusion, we have described photonic crystal guided wave structures tailored for quantum cascade lasers, in which a full TM photonic bandgap is present and lateral electrical injection is possible.  Laser cavity designs in both the mid-IR and THz ranges of the electromagnetic spectrum were considered. We identified candidate defect modes for lasing, and calculated their mode profiles and $Q$-factors.   In addition, we have shown that it is possible - under appropriate conditions - to implement high-contrast photonic structures {\it via} the sole patterning of the top metal layer in THz QC lasers featuring double-metal waveguides.  Such photonic crystal structures are of particular interest for laser structures due to the negligible damage they introduce into the active semiconductor layers and their reduced fabrication complexity.   

\section*{Acknowledgements}
The authors thank F. Julien, C. Sirtori, and Y. Chassagneux for useful discussions.  The work conducted at the Universite Paris-Sud was performed as part of an EURYI scheme award (www.esf.org/euryi), with additional support from the PICS program n.3417.  The Caltech portion of this work was supported by an AFOSR MURI program in plasmonics (http://www.plasmonmuri.caltech.edu/) and a DARPA UPR program in optofluidics (http://www.optofluidics.caltech.edu/).

\bibliography{./biblio}
\end{document}